\documentclass[conference, a4paper]{IEEEtran}

\ifCLASSINFOpdf
\else
\fi

\usepackage{cite}
\usepackage{amsmath,amssymb,amsfonts}

\usepackage{algorithm}
\usepackage{algpseudocode}

\usepackage{graphicx}
\usepackage{textcomp}
\usepackage{xcolor}
\usepackage{siunitx}
\usepackage{import}
\usepackage{tabularx}

\usepackage{multirow}
\usepackage[hidelinks]{hyperref}
\usepackage{tikz,pgfplots}
\usetikzlibrary{matrix}
\usetikzlibrary{patterns}

\newcommand{\RLJSDE}{\mbox{RL-JSDE}}

\usepackage[firstpage=true]{background}
\usepackage{hyperref}

\SetBgContents{\parbox{\textwidth}{\footnotesize © 2022 IEEE. Personal use of this material is permitted. Permission from IEEE must be
		obtained for all other uses, in any current or future media, including
		reprinting/republishing this material for advertising or promotional purposes, creating new
		collective works, for resale or redistribution to servers or lists, or reuse of any copyrighted
		component of this work in other works. DOI: \href{https://doi.org/10.1109/TCSII.2022.3160012}{10.1109/TCSII.2022.3160012.}}}
\SetBgScale{1}
\SetBgAngle{0}
\SetBgPosition{current page.south}
\SetBgVshift{1cm}
\SetBgColor{black}
\SetBgOpacity{1}

\begin{document}
\title{Fast Reconstruction of Three-Quarter Sampling Measurements Using Recurrent Local Joint Sparse Deconvolution and Extrapolation}
\author{\IEEEauthorblockN{Simon~Grosche, Andy~Regensky, Alexander Sinn, J\"urgen~Seiler, and~Andr\'e~Kaup}
	\IEEEauthorblockA{\textit{Multimedia Communications and Signal Processing}\\
		\textit{Friedrich-Alexander University Erlangen-Nürnberg (FAU)}\\
		Cauerstr. 7, 91058 Erlangen, Germany\\
		\{simon.grosche, andy.regensky, alexander.sinn, juergen.seiler, andre.kaup\}@fau.de}
}

\maketitle

\begin{abstract}
Recently, non-regular three-quarter sampling has shown to deliver an increased image quality of image sensors by using differently oriented L-shaped pixels compared to the same number of square pixels. A three-quarter sampling sensor can be understood as a conventional low-resolution sensor where one quadrant of each square pixel is opaque. Subsequent to the measurement, the data can be reconstructed on a regular grid with twice the resolution in both spatial dimensions using an appropriate reconstruction algorithm.
For this reconstruction, local joint sparse deconvolution and extrapolation (L-JSDE) has shown to perform very well. As a disadvantage, L-JSDE requires long computation times of several dozen minutes per megapixel.
In this paper, we propose a faster version of L-JSDE called recurrent L-JSDE (RL-JSDE) which is a reformulation of \mbox{L-JSDE}. For reasonable recurrent measurement patterns, \mbox{RL-JSDE} provides significant speedups on both CPU and GPU without sacrificing image quality. Compared to L-JSDE, \mbox{20-fold} and \mbox{733-fold} speedups are achieved on CPU and GPU, respectively.
\end{abstract}

\IEEEpeerreviewmaketitle

\section{Introduction}
Conventionally, the pixels of an imaging sensor are positioned regularly on the sensor. This results in aliasing whenever higher frequencies than determined by the Nyquist theorem are present in the image.
One solution to circumvent artifacts from aliasing is to employ a non-regular placement of the pixels \cite{Dippe1985, Hennenfent2007, Maeda2009, Anderson2013, Seiler2015, Kovarik2016}. This allows for higher image quality after reconstructing the image on a higher resolution grid without increasing the number of physical pixels compared to a low-resolution sensor.

Aiming for a hardware implementation of non-regular sampling, so called quarter sampling was proposed \cite{Schoberl2011}. For quarter sampling, each pixel of a low-resolution image sensor is covered by 75\% such that only one quadrant of the area of a low-resolution pixel is sensitive to light. The transparent quadrants are placed non-regularly. After a reconstruction of the image on a high resolution grid, even high frequency content can be reconstructed which is otherwise lost using a conventional low-resolution sensor with the same number of pixels. With this, non-regular sampling is able to achieve higher reconstruction qualities than conventional low-resolution sensors by applying appropriate sampling patterns and reconstruction algorithms \cite{Grosche2018, Grosche2021_LFCR}.

\begin{figure}[t]
	\footnotesize
\begingroup%
  \makeatletter%
  \providecommand\color[2][]{%
    \errmessage{(Inkscape) Color is used for the text in Inkscape, but the package 'color.sty' is not loaded}%
    \renewcommand\color[2][]{}%
  }%
  \providecommand\transparent[1]{%
    \errmessage{(Inkscape) Transparency is used (non-zero) for the text in Inkscape, but the package 'transparent.sty' is not loaded}%
    \renewcommand\transparent[1]{}%
  }%
  \providecommand\rotatebox[2]{#2}%
  \newcommand*\fsize{\dimexpr\f@size pt\relax}%
  \newcommand*\lineheight[1]{\fontsize{\fsize}{#1\fsize}\selectfont}%
  \ifx\svgwidth\undefined%
    \setlength{\unitlength}{252bp}%
    \ifx\svgscale\undefined%
      \relax%
    \else%
      \setlength{\unitlength}{\unitlength * \real{\svgscale}}%
    \fi%
  \else%
    \setlength{\unitlength}{\svgwidth}%
  \fi%
  \global\let\svgwidth\undefined%
  \global\let\svgscale\undefined%
  \makeatother%
  \begin{picture}(1,0.34126984)%
    \lineheight{1}%
    \setlength\tabcolsep{0pt}%
    \put(0,0){\includegraphics[width=\unitlength,page=1]{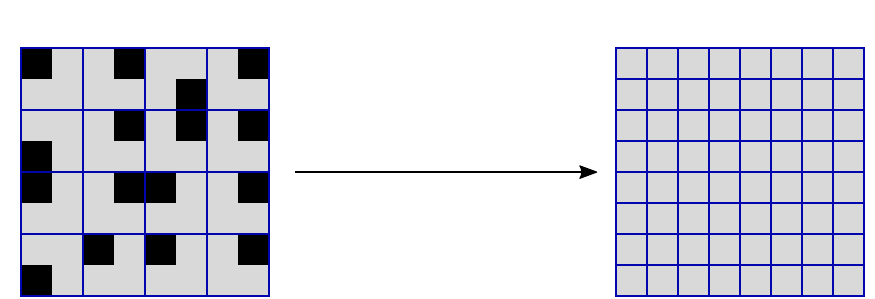}}%
    \put(0.50733105,0.15595792){\color[rgb]{0,0,0}\makebox(0,0)[t]{\lineheight{1.25}\smash{\begin{tabular}[t]{c}Reconstruction\end{tabular}}}}%
    \put(0.02273757,0.30276738){\color[rgb]{0,0,0}\makebox(0,0)[lt]{\lineheight{1.25}\smash{\begin{tabular}[t]{l}Three-quarter sensor\end{tabular}}}}%
    \put(0.70301395,0.30276738){\color[rgb]{0,0,0}\makebox(0,0)[lt]{\lineheight{1.25}\smash{\begin{tabular}[t]{l}Target image\end{tabular}}}}%
  \end{picture}%
\endgroup%

	\normalfont\normalsize
	\caption{Illustration of the concept of Three-quarter sampling. Black regions are opaque, whereas grey regions are transparent. The blue lines separate the individual pixels.}\label{fig:different_pixel_layouts}
\end{figure}
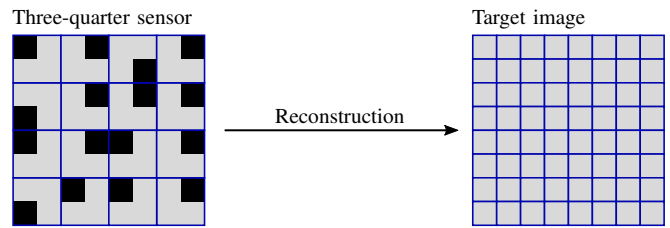

Besides the position of the pixels, also the shape of the pixels can be altered. This led to an advancement called three-quarter sampling  \cite{Seiler2018}. Here, only 25\% of each low-resolution pixel is opaque resulting in L-shaped integration areas. Such arrangement is not only more sensitive to light but also results in higher image qualities after reconstruction compared to quarter sampling. An illustration of three-quarter sampling can be seen in Figure\,\ref{fig:different_pixel_layouts}.

For quarter sampling, an interpolation of the missing pixel on the high resolution grid is required in post-processing. Instead of using simple interpolation techniques such as linear interpolation,  sophisticated algorithms such as the frequency selective reconstruction (FSR) \cite{Seiler2015} had to be developed in order to achieve  competitive image qualities. For three-quarter sampling, the reconstruction is even more involved since the measurement process includes an integration over several high-resolution pixels. Therefore, the so called joint sparse deconvolution and extrapolation (JSDE) \cite{Seiler2018} was proposed. In \cite{Grosche2020_localJSDE}, the JSDE was further generalized to arbitrary local measurements leading to the local JSDE (L-JSDE).
While the reconstruction for three-quarter sampling leads to better image quality, JSDE/L-JSDE also entail a much larger computational complexity compared to FSR.

For this reason, we propose a reformulation of L-JSDE called recurrent L-JSDE (RL-JSDE) in this paper. It enables a much faster processing for the case of \textit{recurrent}, i.e.,  periodically repeating, measurements. As a side effect, using periodically repeating measurements can be considered advantageous  for the hardware manufacturing. Similar assumptions were made in \cite{Grosche2018, Grosche2021_LFCR}. Though we focus on the application of three-quarter sampling, we provide a general description for arbitrary periodically repeating sensor layouts.

The paper is organized as follows: 
In Section\,\ref{sec:compressed_sensing_framework}, we present the concept of three-quarter sampling from a compressed sensing viewpoint. In Section\,\ref{sec:reconstruction_algorithms}, the derivation of L-JSDE is revisited. In Section\,\ref{sec:recurrent_local_JSDE}, the recurrent L-JSDE is proposed.
In Section\,\ref{sec:simulation_and_results}, we describe the performed simulations and evaluate the results in terms of memory requirements and runtime.
\section{Three-Quarter Sampling in the Compressed Sensing Framework}
\label{sec:compressed_sensing_framework}
First, we revisit the definition of a compressed sensing measurement from \cite{Grosche2020_localJSDE}.
Any compressed sensing measurement can be written as a linear combination
\begin{align}\label{eq:yi_from_A_and_f}
	y_i = \sum_{\alpha = 0}^{M-1} \sum_{\beta = 0}^{N-1} A_{i\alpha\beta} f_{\alpha\beta},
\end{align}
where $f_{\alpha\beta}\,{\in}\,[0,1]$ are the gray values of the reference image~$\boldsymbol{f}$ of size $M{\times}N$ that would be acquired with a auxiliary high resolution sensor, $y_i$ are the values measured by the individual pixels of the image sensors and $A_{i\alpha\beta}$ are the coefficients of the image measurement matrix. The index $i\,{\in}\, \{0,\dots, L-1\}$ enumerates the $L$ measurements. Moreover, $\alpha$ and $\beta$ are the vertical and horizontal positions of the pixels of the reference image with the origin positioned in the upper left corner.
In this work, $L\,{=}\,MN/4$.
Other than the vast majority of the compressed sensing literature \cite{Candes2007image, Elad2010, Gan2007, MunFowler2009} that artificially vectorizes the image, we do not vectorize the image in our notation, cf. (\ref{eq:yi_from_A_and_f}).

\begin{figure}[t]
	\footnotesize
	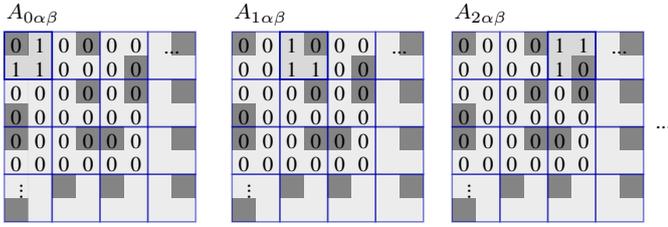
	\normalfont\normalsize
	\caption{First three slices through the measurement matrix superimposed to the corresponding three-quarter sampling sensor.}\label{fig:measurement_matrix}
\end{figure}

Using the notation in (\ref{eq:yi_from_A_and_f}), arbitrary measurement patterns such as quarter sampling \cite{Schoberl2011} and three-quarter sampling \cite{Seiler2018} can be described through the measurement matrix $A_{i\alpha\beta}$.  Figure\,\ref{fig:measurement_matrix} visualizes the first three slices of a \mbox{three-quarter} sampling measurement matrix.

\section{State-of-the-Art Reconstruction Algorithm}
\label{sec:reconstruction_algorithms}

After the measurements have been performed, the image needs to be reconstructed on a regular grid of higher resolution, as already illustrated in Figure\,\ref{fig:different_pixel_layouts}. The reconstruction algorithm has to find an approximate solution $\hat{f}_{\alpha\beta}$ from the sensor measurements $y_i$ such that the measurement equation (\ref{eq:yi_from_A_and_f}) is satisfied as good as possible, e.g., in a least square sense.

For the reconstruction of three-quarter sampling measurements, \mbox{L-JSDE} \cite{Grosche2020_localJSDE} was recently proposed. \mbox{L-JSDE} is a generalization of JSDE \cite{Seiler2018}. It is based on an overlapping sliding window approach and iteratively generates a model in the discrete Fourier transform domain for each model window. We use a target block size of $B{\times}B\,{=}\,4{\times}4$ pixels and a model window of size $W{\times}W\,{=}\,32{\times}32$ pixels. 

The \mbox{L-JSDE} aims at building a model of the image on the model window  $f^\mathrm{local}_{\eta\gamma}$ of size $W{\times}W$. The model, $\hat{f}^\mathrm{local}_{\eta\gamma}$,  can be written as
\begin{align}
	\hat{f}^\mathrm{local}_{\eta\gamma} =  \sum_{\sigma = 0}^{W-1} \sum_{\rho = 0}^{W-1} \Phi_{\eta\gamma \sigma\rho}  \hat{c}_{\sigma\rho},
\end{align} 
using a sparse transform such as the Fourier transform $\Phi_{\eta\gamma \sigma\rho} = \mathrm{e}^{2\pi \mathrm{j} \frac{\eta \sigma}{M}} \mathrm{e}^{2\pi \mathrm{j} \frac{\gamma \rho}{N}}$. Most coefficients $\hat{c}_{\sigma\rho}$ are close or equal to zero since the image is approximately sparse in the Fourier domain \cite{Lam2000, Elad2010}. 
L-JSDE changes one coefficient $\hat{c}_{\sigma\rho}$ at a time. The selected frequency  component $(\sigma,\rho)$ as well the amount of the change are determined optimally with respect to the measurement error in every iterations. For more details and a complete derivation, see \cite{Grosche2020_localJSDE}.
Though the model is built on the entire model window, \mbox{L-JSDE} only uses the reconstructed target block of size $B{\times}B$ for the final image. Afterwards, the next target block is reconstructed in the same manner.

\section{Proposed Recurrent \mbox{L-JSDE}}
\label{sec:recurrent_local_JSDE}
In this section, we propose recurrent L-JSDE (RL-JSDE) being a novel reformulation of \mbox{L-JSDE} \cite{Grosche2020_localJSDE}. It is useful in case the sensor layout is recurrent such that the measurement matrix is identical for many target blocks during the reconstruction.
The concepts used for RL-JSDE are related to an approach presented in \cite{Seiler2011} where selective extrapolation \cite{Kaup2005}, being an inpainting algorithm and the predecessor of FSR \cite{Seiler2015}, is sped up by several pre-computations for the case of arbitrary basis functions.

In the remainder of this section, we present all steps to derive the RL-JSDE.
As the L-JSDE, the \mbox{RL-JSDE} aims at building a model of the image in the model window  $f^\mathrm{local}_{\eta\gamma}$ of size $W{\times}W$. The model, $\hat{f}^\mathrm{local}_{\eta\gamma}$,  can be written as
\begin{align}
	\hat{f}^\mathrm{local}_{\eta\gamma} =  \sum_{\sigma = 0}^{W-1} \sum_{\rho = 0}^{W-1} \Phi_{\eta\gamma \sigma\rho}  \hat{c}_{\sigma\rho}.
\end{align}

The residual signal can now be written as
\begin{align}
	r_m = y^\mathrm{local}_m - \sum_{\eta = 0}^{W-1}\sum_{\gamma = 0}^{W-1}  A^\mathrm{local}_{m\eta\gamma} \hat{f}^\mathrm{local}_{\eta\gamma},
\end{align}
being a measure of how closely the model reproduces the measurement $y^\mathrm{local}_m$.

Starting with a model $\hat{f}^{\mathrm{local}, (\nu)}=0$ as initialization, one coefficient of the model is changed in each iteration. Let us assume that in the $\nu$-th iteration, the coefficient $\hat{c}_{\sigma\rho}$ is updated by $\delta^{(\nu)}_{\sigma\rho}$,
\begin{align}
	\hat{c}^{(\nu)}_{\sigma\rho} = \hat{c}^{(\nu-1)}_{\sigma\rho} + \delta^{(\nu)}_{\sigma\rho} ,
\end{align}
where the indices $\sigma$ and $\rho$ are only written out for easier remembrance.
This implies that the model is updated to
\begin{align}
	\hat{f}^{\mathrm{local}, (\nu)}_{\eta\gamma} = \hat{f}^{\mathrm{local},(\nu-1)}_{\eta\gamma} + \delta^{(\nu)}_{\sigma\rho} \Phi_{\eta\gamma \sigma\rho},
\end{align}
and the residual signal can therefore be written as
\begin{align}
	r_{m}^{(\nu)} = r_m^{(\nu-1)} - \delta^{(\nu)}_{\sigma\rho} \sum_{\eta,\gamma = 0}^{W-1}  A^\mathrm{local}_{m\eta\gamma} \Phi_{\eta\gamma \sigma\rho}.
	\label{eq:r_m_nu_from_prev_step}
\end{align}
The weighted residual energy in the $\nu$-th step can now be expressed as
\begin{align}
	E_{\mathrm{w}}^{(\nu)} &= \sum_{m=0} \left|r_m^{(\nu)}\right|^2  w_m,  %
	\label{eq:E_weight_nu}
\end{align}
where $w_m$ is the isotropic spatial weighting function that gives less weight to the errors further away from the center in order to optimize to model inside the target block \cite{Grosche2020_localJSDE}.
The coefficient update $\delta^{(\nu)}_{\sigma\rho}$ shall be chosen such that the weighted residual energy $E_{\mathrm{w}}^{(\nu)}$ is minimized. For this, Wirtinger calculus is used and the partial derivatives with respect to the coefficient updates are set to zero,
\begin{align}
	\frac{\partial E_{\mathrm{w}}^{(\nu)}}{\partial \delta^{(\nu)}_{\sigma\rho}}
	\overset{!}{=} 0  \quad \mathrm{and} \quad   
	\frac{\partial E_{\mathrm{w}}^{(\nu)}}{\partial \left(\delta^{(\nu)}_{\sigma\rho}\right)^*} \overset{!}{=} 0.
\end{align}

Fortunately, these derivatives can be evaluated analytically. After straightforward but cumbersome calculations, the authors of \cite{Grosche2020_localJSDE} arrive at the update equations in the  $\nu$-th iteration,
\begin{align}
	(u, v) &= \underset{\sigma, \rho}{\mathrm{argmin}}\, \left(q_{\sigma\rho} \cdot \left(E_{\mathrm{w}}^{(\nu)}-E_{\mathrm{w}}^{(\nu-1)}\right)\right) = \ldots =\nonumber\\
	&\hspace{-1cm}=
	\underset{\sigma,\rho}{\mathrm{argmax}}\, \left(
	q_{\sigma\rho}\frac{
		\left|\sum\limits_{m=0} \sum\limits_{\tilde\eta,\tilde\gamma = 0}^{W-1} A^\mathrm{local}_{m\tilde\eta\tilde\gamma} \Phi^*_{\tilde\eta\tilde\gamma \sigma\rho} w_m r_m^{(\nu-1)}\right|^2
	}{
		\sum\limits_{m=0}  w_m \left|\sum\limits_{\eta\gamma = 0}^{W-1} \sum\limits_{\gamma = 0}^{W-1} A^\mathrm{local}_{m\eta\gamma} \Phi_{\eta\gamma \sigma\rho}\right|^2
	} \right)
	\label{eq:argmax_final_JSDE}
\end{align}

Here, we identify that most calculations in the numerator and the denominator could be pre-computed in case of recurrent measurements. Therefore, we rewrite (\ref{eq:argmax_final_JSDE}) as
\begin{align}\label{eq:argmax_uv}
	(u,v) = \underset{\sigma,\rho}{\mathrm{argmax}}\,\left( q_{\sigma\rho} \frac{\left|R^{(\nu-1)}_{\sigma\rho}\right|^2}{D_{\sigma\rho}} \right).
\end{align}
Here, $R^{(\nu)}_{\sigma\rho}$ is the projected residual 
\begin{align}
	R^{(\nu)}_{\sigma\rho} = \sum\limits_{m=0} \underbrace{\sum\limits_{\tilde\eta,\tilde\gamma = 0}^{W-1} A^\mathrm{local}_{m\tilde\eta\tilde\gamma} \Phi^*_{\tilde\eta\tilde\gamma \sigma\rho} w_m}_{=: B_{m\sigma\rho}} r_m^{(\nu)},
	\label{eq:projected_residual_def}
\end{align}
with $B$ being the projection matrix. $D$ is an auxiliary matrix defined as a slice 
\begin{align}
	D_{\sigma\rho}:=C_{\sigma\rho\sigma\rho},
\end{align}
through the four-dimensional matrix $C$ with coefficients
\begin{align}\label{eq:C_lookup}
	C_{\sigma\rho uv} := \sum\limits_{m=0} \sum\limits_{\tilde\eta,\tilde\gamma = 0}^{W-1} \sum_{\eta, \gamma = 0}^{W-1} A^\mathrm{local}_{m\tilde\eta\tilde\gamma} \Phi^*_{\tilde\eta\tilde\gamma \sigma\rho} w_m   A^\mathrm{local}_{m\eta\gamma} \Phi_{\eta\gamma uv},
\end{align}
and $q_{\sigma\rho}$ is the frequency weighting function that prefers low frequencies over high frequencies.
The auxiliary matrices $B$, $C$, and $D$ only depend on the local measurement matrix, the Fourier basis functions and the spatial weighting function $w_m$ and can therefore be pre-computed and re-used for all model windows with the same measurement matrix.

With the optimal frequency components $(u,v)$ at hand, we update the model coefficient 
\begin{align}
	\hat{c}^{(\nu)}_{uv} = \hat{c}^{(\nu-1)}_{uv} + \gamma_\mathrm{odc} \delta^{(\nu)}_{uv},
\end{align}
using the expansion coefficient
\begin{align}\label{eq:expan_coeff_update}
	\delta^{(\nu)}_{uv} = \frac{R^{(\nu-1)}_{uv}}{D_{uv}},
\end{align}
and the so called orthogonality deficiency compensation factor $\gamma_\mathrm{odc}$, which was first introduced in \cite{Seiler2008} and can be interpreted as a step-width.
All coefficients of the projected residual are updated according to
\begin{align}\label{eq:proj_res_update}
	R^{(\nu)}_{\sigma\rho} = R^{(\nu-1)}_{\sigma\rho} - \gamma_\mathrm{odc} \delta^{(\nu)}_{uv}  C_{\sigma\rho uv}.
\end{align}
After the maximum number of iterations, $\nu_\mathrm{max}$, we transform the model back to the image domain
\begin{align}
	\hat{f}^\mathrm{local}_{\eta\gamma} =  \sum_{\sigma = 0}^{W-1} \sum_{\rho = 0}^{W-1} \Phi_{\eta\gamma \sigma\rho}  \hat{c}^{(\nu = \nu_\mathrm{max})}_{\sigma\rho}
\end{align}
and copy the central target block of size $B{\times}B$ pixels to the final image. Afterwards, the next target block is processed.

\begin{figure}[t]
	\footnotesize
	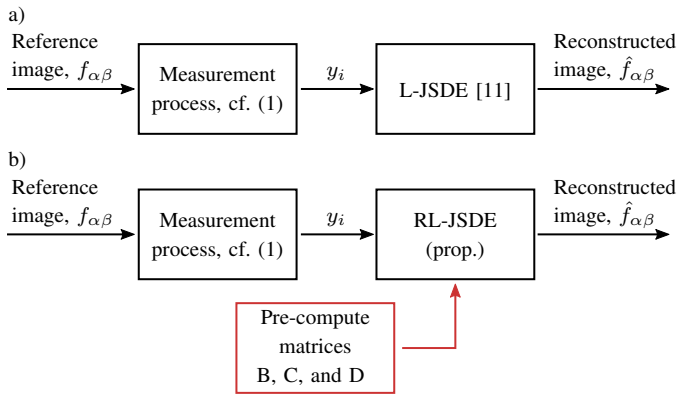
	\normalfont\normalsize
	\caption{Flow chart for (a) L-JSDE and (b) RL-JSDE. For RL-JSDE, several matrices are pre-computed under the assumption that the measurements are recurrent.}\label{fig:flowchart_rljsde}
\end{figure}

The reconstruction result of \RLJSDE{} is identical to that of \mbox{L-JSDE} up to numerical precision. Comparing the above steps with the derivation of \mbox{L-JSDE} \cite{Grosche2020_localJSDE}, the key difference is that the newly defined projected residual  can directly be updated in (\ref{eq:proj_res_update}) instead of having to update the residual itself.  The projected residual can then be used in (\ref{eq:argmax_uv}) and (\ref{eq:expan_coeff_update}) directly and the residual signal itself is not needed. Moreover, the matrices $B$, $C$, and $D$ are identified and pre-computed for any of the $W^2/B^2$ measurement matrices needed during the reconstruction. A flow-chart comparing L-JSDE and RL-JSDE is shown in Figure\,\ref{fig:flowchart_rljsde}.

Overall, the computational complexity for each target block scales linearly with the number of pixels inside the model window, i.e., $\mathcal{O}(W^2)$. Other than this, for L-JSDE, the complexity scales with $\mathcal{O}(W^4)$ as can be seen in (\ref{eq:argmax_final_JSDE}) where an additional matrix-vector product has to be computed.

\section{Experiments and results}
\label{sec:simulation_and_results}
\subsection{Simulation setup}
Several experiments have been performed to evaluate the reconstruction quality, the memory requirements and the runtime of RL-JSDE compared to L-JSDE.
The code of L-JSDE was provided by the authors and we implemented RL-JSDE within the same framework.\footnote{The source code of RL-JSDE is published online:\\\texttt{https://gitlab.lms.tf.fau.de/LMS/local\_jsde\_public}}
Regarding the evaluation dataset, we use images from the TECNICK dataset \cite{Asuni2014} consisting of 100 natural images of size $1200{\times}1200$ pixels. 
The images serve as reference images $\boldsymbol{f}$ and the measured values $y_i$ can be generated by multiplying the image with the measurement matrix of the three-quarter sampling sensor  as in (\ref{eq:yi_from_A_and_f}). For the recurrent sensor layout, a random pattern is used that repeats periodically after 32 pixels.

The reconstruction results of L-JSDE and  RL-JSDE were verified to be identical up to numerical precision for all images, as expected. In the following, we focus on the additional memory requirements and the runtimes of the different reconstruction algorithms.

\subsection{Additional Memory Usage}
The RL-JSDE performs several pre-computations leading to an additional need of memory compared to L-JSDE.
\begin{table}[t]
	\caption{Additional memory required by RL-JSDE for all matrices $B, C,$ and $D$ due to the pre-computations.}
	\label{tab:memory}
	
	\centering
	\setlength{\tabcolsep}{3pt}
	\begin{tabularx}{.85\columnwidth}{l||c|c|c||c}
		 & $B$ & $C$ & $D$  & Total\\\hline
		Required memory            & 134.4 MB & 537.6 MB & 0.5MB & 672.5MB
	\end{tabularx}
\end{table}
In Table\,\ref{tab:memory} the required memory to store all $32^2/4^2 = 64$ matrices $B, C,$ and $D$ is given. The total used memory to  store the pre-computations is less than 700MB being acceptable on modern computers and modern GPUs.
The memory requirements beyond the pre-computations are on the order of kilobytes and are therefore negligible.

\subsection{Evaluation of the Runtime}
In this section, we investigate the runtimes for the different reconstruction algorithms. 
For the CPU measurements, all algorithms were restricted to a single CPU core of an \textit{Intel Xeon E3-1245v5} processor with 3.50\,GHz. In each case, the measured runtime was averaged for all 100 images of the TECNICK dataset being of size $1200{\times}1200$ pixels.
For the GPU measurements, we rewrote both algorithms using the Numba \cite{Numba2015} GPU interface and ran the reconstructions on an \textit{Nvidia RTX 2080}.

\begin{table}[t]
	\caption{Average runtimes in seconds for an image of size $1200{\times}1200$ pixels.}
	\label{tab:results_runtime}
	
	\centering
	\setlength{\tabcolsep}{3pt}
	\begin{tabularx}{.8\columnwidth}{l||c|c}
		Reconstruction algorithm &  L-JSDE \cite{Grosche2020_localJSDE} & RL-JSDE (prop.) \\\hline
		Runtime  on CPU              & $1823$\,s& $86$\,s\\
		Runtime on GPU & $550$\,s & $0.75$\,s
	\end{tabularx}
\end{table}

The timing results are provided in Table\,\ref{tab:results_runtime}. We find that the proposed \RLJSDE{} is more than 20x faster than L-JSDE \cite{Grosche2020_localJSDE} on the CPU. For the GPU, a 733-fold speedup is achieved. 
Compared to the CPU version of L-JSDE \cite{Grosche2020_localJSDE}, the GPU version of RL-JSDE achieves a 2430-fold speedup. With this, near real-time processing for the reconstruction of recurrent three-quarter sampling measurements seems possible.

\section{Conclusion and Future Work}
\label{sec:conclusion}
In this paper, we proposed a reformulation of L-JSDE called recurrent L-JSDE (RL-JSDE). With this reformulation, several pre-calculations can be done for the case of recurrent sensor layouts and the complexity of the algorithm can be reduced. As exemplarily recurrent sensor layout, we use a recurrent three-quarter sampling sensor design.

In our evaluation, we achieve a 20-fold speedup on the CPU and a 733-fold speedup on the GPU. Overall, more than one image with $1200{\times}1200$ pixels could be reconstructed per second. At the same time, the reconstruction results of \RLJSDE{} are identical to that of \mbox{L-JSDE} up to numerical precision.

With such significant speed-ups in reconstruction, this work paves the way to realize three-quarter sampling sensors in actual hardware systems simplifying both sensor manufacturing through recurrent measurement patterns and enabling fast image reconstruction through RL-JSDE.

\ifCLASSOPTIONcaptionsoff
  \newpage
\fi

\bibliographystyle{IEEEtran}
\bibliography{literatur_jabref}

\end{document}